\documentclass[aps,pra,showpacs,twocolumn]{revtex4}

\usepackage{dcolumn}
\usepackage{bm}
\usepackage{amsmath}
\usepackage{graphicx}

\usepackage{amsmath,amssymb}

\newcommand{\mean}[1]{\langle{#1}\rangle}

\newcommand{\bra}[1]{\langle{#1}|}
\newcommand{\ket}[1]{|{#1}\rangle}

\newcommand{\dgg}{^{\dagger}}

\newcommand{\Tr}{{\rm Tr}\hspace{0.07cm}}

\begin{document}



\title{Dissipation-induced pure Gaussian state}



\author{Kei Koga}
\author{Naoki Yamamoto}
\email[]{yamamoto@appi.keio.ac.jp}


\affiliation{
Department of Applied Physics and Physico-Informatics, 
Keio University, 
Yokohama 223-8522, 
Japan}


\date{\today}


\begin{abstract} 

This paper provides some necessary and sufficient conditions 
for a general Markovian Gaussian master equation to have a 
unique pure steady state. 
The conditions are described by simple matrix equations, thus 
the so-called environment engineering problem for pure Gaussian 
state preparation can be straightforwardly dealt with in the 
linear algebraic framework. 
In fact, based on one of those conditions, for an arbitrary 
given pure Gaussian state, we obtain a complete parameterization 
of the Gaussian master equation having that state as a unique 
steady state; 
this leads to a systematic procedure for engineering a desired 
dissipative system. 
We demonstrate some examples including Gaussian cluster states. 

\end{abstract}





\maketitle



\section{Introduction}

Preparing a desired pure state, particularly under influence 
of dissipation, is clearly a most important subject in 
quantum information technologies. 
To tackle this problem, other than some well-acknowledged 
strategies such as quantum error correction, recently we have 
a totally different method that rather utilizes dissipation 
\cite{Poyatos,Cirac1,Kraus,Kraus_nature,TicozziViola2008,
Ficek2009,TicozziViola2011,VerstraeteNaturePhys,Vollbrecht,
Schirmer,Cirac2,Polzik,Yamamoto}. 
The basic idea is to engineer a dissipative system so that 
the system state $\hat\rho(t)$ governed by the corresponding 
Markovian master equation
\begin{equation}
\label{me}
   \frac{d\hat \rho}{dt} 
    = - i[\hat H , \hat\rho] 
      + \sum_{k=1}^m \Big( 
           \hat L_k \hat\rho \hat L_k^\dagger 
             -\frac{1}{2} \hat L_k^\dagger \hat L_k \hat\rho 
               -\frac{1}{2} \hat\rho \hat L_k^\dagger \hat L_k \Big)
\end{equation}
{\it must} evolve towards a desired pure state: 
$\hat\rho(t)\rightarrow\ket{\phi}\bra{\phi}$ as $t\rightarrow\infty$. 
Here, $\hat H$ is the system Hamiltonian and $\hat L_k$ 
$(k=1,\ldots,m)$ is the dissipative channel that represents the 
coupling between the system and the $k$-th environment mode. 
The main advantage of this {\it environment engineering} approach 
is that the dissipation-induced state $\ket{\phi}$ is robust 
against any perturbation and thus may serve as a desired state, 
e.g., an entangled state for quantum computation 
\cite{VerstraeteNaturePhys} and quantum repeater \cite{Vollbrecht}. 
Therefore, a complete characterization of the master equation 
having a unique pure steady state should be of great use, and 
actually in the finite-dimensional case it was given by 
Kraus et al. \cite{Kraus}. 
In particular, they showed that some useful pure states 
including cluster states can be prepared by {\it quasi-local} 
dissipative process, i.e., dissipative channels that act only 
on a small number of subsystems.

In this research direction the infinite-dimensional counterpart 
of the above-mentioned state preparation method should be explored. 
In particular, Gaussian states constitute a wide and important 
class of quantum states, which serve as the basis for various 
continuous-variable (CV) quantum information processing 
\cite{Braunstein2005,Furusawa2011}. 
The contribution of this paper is to provide some necessary 
and sufficient conditions for the master equation \eqref{me} to 
have a unique pure steady state, when $\hat H$ and $\hat L_k$ 
are of general form for $\hat\rho(t)$ to be Gaussian for all $t$. 
The conditions are described by simple matrix equations, thus 
they can be properly applied to the above-mentioned environment 
engineering problem for pure Gaussian state preparation. 
Actually, one of those conditions enables us to obtain a 
complete parameterization of the Gaussian master equation 
that uniquely has a pure steady state. 
This leads to a systematic procedure for constructing a dissipative 
system deterministically yielding a desired pure Gaussian state. 
We provide some examples of dissipation-induced states including 
the so-called Gaussian cluster states \cite{Zhang2006,vanLoock2007,
Menicucci2011}, which are known as essential resources for the 
CV one-way quantum computing \cite{Menicucci2006}, with focusing 
on how they can actually be prepared by quasi-local dissipative 
process.


\section{Gaussian dissipative systems}

We here provide the phase space representation of the general 
Gaussian dissipative system with $n$-degrees of freedom, which 
is subjected to the master equation \eqref{me}. 
Let $(\hat q_i,~\hat p_i)$ be the canonical conjugate pair of 
the $i$-th subsystem. 
It then follows from the canonical commutation relation 
$[\hat q_i,~\hat p_j]=i \delta _{ij}$ that 
the vector of system variables 
$\hat x :=(\hat q_1, \ldots, \hat q_n, \hat p_1, \ldots, \hat p_n)^\top$ 
satisfies 
\begin{eqnarray*}
   \hat x \hat x ^\top -(\hat x \hat x^\top )^\top 
      = i \Sigma,~~
   \Sigma = \left(\begin{array}{cc}
               0 & I_n \\
               -I_n & 0
            \end{array}\right), 
\end{eqnarray*}
where $I_n$ denotes the $n\times n$ identity matrix and $\top$ 
the matrix transpose. 
A Gaussian state is completely characterized by only the mean 
vector $\mean{\hat x}$ and the (symmetrized) covariance matrix 
$V=\mean{\Delta \hat x \Delta \hat x^\top 
 + (\Delta \hat x \Delta \hat x^\top)^\top}/2,~
\Delta \hat x=\hat x-\mean{\hat x}$. 
Here the mean $\mean{\hat X}=\Tr(\hat X\hat \rho)$, with $\hat \rho$ 
the corresponding Gaussian density operator, is taken elementwise; 
e.g., $\mean{\hat x}=(\mean{\hat q_1}, \ldots, \mean{\hat p_n})^\top$. 
The uncertainty relation is represented by the matrix inequality 
$V+i\Sigma/2 \geq 0$, thus $V>0$ \cite{StefanoLloyd}. 
Note also that the purity of a Gaussian state is simply given by 
$P=\Tr(\hat \rho^2)=1/\sqrt{2^{2n}{\rm det}(V)}$. 
Now consider Eq.~\eqref{me} with $\hat H$ and $\hat L_k$ given by
\begin{equation*}
     \hat H = \frac{1}{2} \hat x^\top G \hat x,~~
     \hat L_k = c_k^\top \hat x, 
\end{equation*}
where $G=G^\top \in \textbf{R}^{2n \times 2n}$ and 
$c_k \in \textbf{C}^{2n}$ are the parameter matrix and vector 
specifying the dissipative system. 
Then, the time-evolution of $\mean{\hat x}$ and $V$ read 
\begin{equation}
\label{statistics}
  \frac{d\mean{\hat x}}{dt} = A \mean{\hat x},~~
  \frac{dV}{dt} = AV+VA^\top + D, 
\end{equation}
where $A=\Sigma[G+\Im(C^\dagger C)]$ and 
$D=\Sigma\Re(C^\dagger C)\Sigma^\top$ with 
$C=(c_1,\ldots,c_m)^\top\in \textbf{C}^{m\times 2n}$ 
($\Re$ and $\Im$ denote the real and imaginary parts, 
respectively). 
If $\hat\rho(0)$ is Gaussian, then $\hat\rho(t)$ is always 
Gaussian with mean $\mean{\hat x(t)}$ and covariance matrix 
$V(t)$. 
See \cite{Wiseman} for detailed description.

Now suppose that $A$ is {\it Hurwitz}; 
i.e., all the eigenvalues of $A$ have negative real parts. 
This is equivalent to that the system has a unique steady state; 
it is Gaussian with mean $\mean{\hat x(\infty)}=0$ and covariance 
matrix $V_s$ that is a unique solution to the following 
matrix equation: 
\begin{eqnarray}
\label{ale}
   AV_s + V_s A^\top + D = 0. 
\end{eqnarray} 
Note that $V_s$ can be explicitly represented as
\begin{eqnarray}
\label{steady solution}
    V_s=\int_0^\infty e^{At} D e^{A^\top t} dt. 
\end{eqnarray}
The purpose of this paper is, as mentioned before, to fully 
characterize a Gaussian master equation that has a unique pure 
steady state, and this is now expressed in terms of 
Eq. \eqref{steady solution} by $2^{2n}{\rm det}(V_s)=1$. 
However, clearly this is not useful. 
In the next section we give a much simpler and explicit 
version of such a characterization.


\section{The dissipation-induced pure Gaussian states}


\subsection{Pure steady state condition}

Here we address our first main result:

{\it Theorem 1}: 
Suppose that Eq. \eqref{ale} has a unique solution $V_s$. 
Then the following three conditions are equivalent: 
\begin{itemize}
\setlength{\itemsep}{0pt}
\item[(i)]
The system has a unique pure steady state with covariance 
matrix $V_s$. 
\item[(ii)]
$V_s$ satisfies the following matrix equations:
\begin{eqnarray}
& & \hspace*{0em}
\label{th1-1}
        \Big(V_s + \frac{i}{2}\Sigma \Big)C^\top = 0, 
\\ & & \hspace*{0em}
\label{th1-2}
        \Sigma G V_s + V_s G \Sigma^\top =0.
\end{eqnarray}
\item[(iii)]
The following matrix equation holds: 
\begin{equation}
\label{ksigmac}
     K \Sigma C^\top = 0, 
\end{equation}
where $K\in{\bf C}^{2nm\times 2n}$ is defined by
\begin{equation}
\label{Kc}
    K = (C^\top, \ G\Sigma^\top C^\top, \cdots, 
          (G \Sigma^\top)^{2n-1}C^\top )^\top. 
\end{equation}
\end{itemize}
Furthermore, when the above equivalent conditions are satisfied, 
$V_s$ is represented by 
\begin{equation}
\label{purev}
     V_s=\frac{1}{2}\Sigma^\top \Im(K^\dagger K) 
            [\Re (K^\dagger K)]^{-1}. 
\end{equation}

To prove this theorem we need the following lemma.

{\it Lemma 1}: 
Suppose that Eq. \eqref{ale} has a unique solution. 
Then ${\rm rank}(\bar{K}^\top)=2n$, 
where 
\begin{equation*}
    \bar{K} = (\bar C^\top,~ G\Sigma^\top \bar C^\top,\ldots, 
                (G \Sigma^\top )^{2n-1} \bar C^\top )^\top
            \in{\bf R}^{4nm\times 2n}
\end{equation*}
with 
$\bar C = (\Re(C)^\top, \Im(C)^\top)^\top
\in{\bf R}^{2m\times 2n}$.

{\it Proof of Lemma 1}: 
From the assumption, Eq. \eqref{ale} has a unique solution 
\eqref{steady solution}. 
Suppose there exists a vector $\xi\in{\bf R}^{2n}$ such that 
$\bar{K} \xi = 0$. 
Then, noting that $A$ and $D$ are written by 
$A = \Sigma [G + \bar C^\top \Sigma\bar C]$ and 
$D = \Sigma \bar C^\top \bar C \Sigma^\top$, we have 
$\xi^\top \Sigma^\top V_s \Sigma \xi 
= \int_0^\infty \|\bar C \Sigma^\top e^{A^\top t} \Sigma \xi\|^2 dt=0$, 
and this is contradiction to $V_s>0$. 
Thus ${\rm rank}(\bar{K}^\top)=2n$.  
\hfill $\blacksquare$


{\it Proof of Theorem 1}: 
The proof is divided into four steps.

\mbox{1. (i) $\Rightarrow$ (ii)}. 
From the Williamson theorem \cite{Williamson} a covariance 
matrix $V$ corresponding to a pure state is expressed by 
$V=SS^\top/2$ with $S$ a symplectic matrix \cite{Simon}. 
Thus substituting $V_s=SS^\top/2$ for Eq. \eqref{ale}, 
multiplying it by $S^{-1}$ from the left and by $S^{-\top}$ 
from the right, and finally using this equation twice to erase $G$, 
we have 
\begin{eqnarray}
& & \hspace*{-2em}
    (C_r'^\top - \Sigma C_i'^\top)
       (C_r'^\top - \Sigma C_i'^\top)^\top 
\nonumber \\ & & \hspace*{0em}
    \mbox{}
    +(C_i'^\top + \Sigma C_r'^\top)
       (C_i'^\top + \Sigma C_r'^\top)^\top = 0, 
\nonumber
\end{eqnarray}
where $C_r'=\Re(C)S$ and $C_i'=\Im(C)S$. 
Note $S\Sigma S^\top = \Sigma$. 
Thus 
$C_r'^\top - \Sigma C_i'^\top = C_i'^\top + \Sigma C_r'^\top = 0$, 
which immediately leads to 
$V_s \Re(C)^\top = \Sigma \Im(C)^\top/2$ and 
$V_s \Im(C)^\top = -\Sigma \Re(C)^\top/2$. 
From these equations we obtain Eq. \eqref{th1-1}. 
Combining this with Eq. \eqref{ale} yields Eq. \eqref{th1-2}.

\mbox{2. (ii) $\Rightarrow$ (iii)}. 
Multiplying Eq.~\eqref{th1-1} by $\Sigma G$ from the left and 
using Eq. \eqref{th1-2}, we get 
$(V_s + i\Sigma/2)G\Sigma^\top C^\top = 0$. 
Repeating this manipulation yields 
\begin{equation}
\label{uncertaintyc}
   (V_s + i\Sigma/2)K^\top = 0, 
\end{equation}
with $K$ defined in Eq. \eqref{Kc}. 
This readily leads to $K(V_s \pm i\Sigma/2)K^\top = 0$, 
thus $K\Sigma K^\top = 0$. 
The Cayley-Hamilton theorem implies Eq. \eqref{ksigmac}.

\mbox{3. Derivation of Eq. \eqref{purev}}. 
Define the $4nm\times 2n$ matrix 
$K'=(\Re(K^\top),~\Im(K^\top))^\top$, 
then the real and imaginary parts of Eq. \eqref{uncertaintyc} are 
summarized in a single equation as 
\[
    \Sigma V_s K'^\top 
      = K'^\top \Sigma/2. 
\]
Now, from the assumption, we can use Lemma~1 and find 
${\rm rank}(K'^\top)={\rm rank}(\bar{K}^\top)=2n$. 
Then multiplying the above equation by $K'(K'^\top K')^{-1}$ 
(the generalized inverse matrix of $K'^\top$) from the right, 
we have 
\[
   \Sigma V_s = (K_r^\top K_i - K_i^\top K_r )
                  (K_r^\top K_r + K_i^\top K_i)^{-1}/2,
\]
where $K_r=\Re(K)$ and $K_i=\Im(K)$. 
This is just Eq.~(\ref{purev}).

Now let us verify that Eq.~\eqref{purev} is symmetric. 
Noting that Eq. \eqref{ksigmac} is equivalent to 
$K \Sigma K^\top = 0$, we have 
\begin{equation}
\label{sym1}
    \left(\begin{array}{c}
      K \\
      K^*
    \end{array}\right)
       \Sigma ( K^\top,~-K^\dagger )
   =\left(\begin{array}{c}
      K \\
      -K^*
    \end{array}\right)
       \Sigma^\top ( K^\top,~K^\dagger ). 
\end{equation}
Then multiplying this equation by $(K^\dagger,~K^\top)$ from 
the left and by $(K^\dagger,~K^\top)^\top$ from the right, 
we obtain 
\[
   \Re(K^\dagger K) \Sigma \Im(K^\dagger K) 
     = -\Im(K^\dagger K) \Sigma^\top \Re(K^\dagger K), 
\]
thus $V_s=V_s^\top$.

\mbox{4. (iii) $\Rightarrow$ (i)}. 
First, to show that the state is pure, we use the fact \cite{Wolf} 
that, for a pure Gaussian state, the corresponding covariance matrix 
$V$ satisfies 
\begin{equation}
\label{purecondition}
    \Sigma V \Sigma V = -I/4. 
\end{equation}
Now multiply Eq. \eqref{sym1} by $(K^\dagger,~K^\top)$ 
from the left and by $(K^\dagger,~-K^\top)^\top$ from the right, 
then we have 
\[
    \Re(K^\dagger K) \Sigma \Re (K^\dagger K) 
     = \Im(K^\dagger K) \Sigma^\top \Im(K^\dagger K). 
\]
This equation readily implies that $V_s$ given by Eq. \eqref{purev} 
satisfies Eq. \eqref{purecondition}, hence the corresponding state 
is pure.

Next, let us show that Eq. \eqref{purev} satisfies Eq. \eqref{ale}. 
Note that Eq. \eqref{purev} is rewritten as 
$V_s\Re(K^\dagger K) = - \Sigma \Im(K^\dagger K)/2$. 
Also from Eq. \eqref{purecondition} we have 
$V_s \Im(K^\dagger K) = \Sigma \Re(K^\dagger K)/2$. 
These two equations yield 
\[
   (V_s -i\Sigma/2)K^\dagger K = 0, 
\]
which is equivalent to $(V_s - i\Sigma/2) K^\dagger = 0$, 
thus Eq.~\eqref{uncertaintyc}. 
This implies that Eq. \eqref{th1-1} holds. 
Moreover, Eq. \eqref{uncertaintyc} leads to 
$(V_s G\Sigma^\top +i\Sigma G \Sigma^\top/2)K^\top = 0$ and 
$(\Sigma G V_s + i\Sigma G \Sigma/2)K^\top = 0$; 
from these equations we have 
$(\Sigma G V_s + V_s G \Sigma^\top)K^\top = 0$, and as now 
$\Sigma G V_s + V_s G \Sigma^\top$ is real and 
${\rm rank}(K'^\top)= 2n$, we obtain Eq. \eqref{th1-2}. 
As a result, $V_s$ satisfies Eqs. \eqref{th1-1} and \eqref{th1-2}, 
but these two equations correspond to a specific decomposition of 
Eq. \eqref{ale}. 
That is, $V_s$ is the solution to Eq. \eqref{ale}. 
\hfill $\blacksquare$


We give an interpretation of Theorem~1. 
Let $\hat \rho_{s}$ be the pure density operator corresponding 
to the covariance matrix $V_s$. 
Then Eqs. \eqref{th1-1} and \eqref{th1-2} are equivalent to 
\begin{equation}
\label{dark state condition}
\hspace{-0em}
   2\hat L_k\hat\rho_s\hat L_k\dgg
      -\hat L_k\dgg\hat L_k\hat\rho_s
          -\hat\rho_s\hat L_k\dgg\hat L_k = 0~~\forall k,~~~~
   [\hat H, \hat \rho_s]=0, 
\end{equation}
respectively. 
The former condition is further equivalent to that 
$\hat L_k\ket{\phi_s}$ is parallel to $\ket{\phi_s}$ for all $k$, 
where $\hat\rho_s=\ket{\phi_s}\bra{\phi_s}$. 
This means that $\hat\rho_s$ is the so-called {\it dark state}; 
that is, Eqs. \eqref{th1-1} and \eqref{th1-2} are the phase space 
representation of the condition for the state to be dark. 
Moreover, the uniqueness of $\hat\rho_s$ allows us to erase itself 
in Eq. \eqref{dark state condition} and obtain a single equation 
with respect to $\hat H$ and $\hat L_k$. 
The phase space representation of this equation is no more than 
Eq.~\eqref{ksigmac}. 
It should be maintained that $\hat\rho_s$ is explicitly 
represented in a directly computable form \eqref{purev}.


Next let us discuss how to use Theorem~1, particularly for the 
purpose of environment engineering. 
First, note that Eq.~\eqref{ksigmac} depends only on the system 
matrices $G$ and $C$; thus the condition (iii) should be applied, 
for a given specific system configuration, to find the system 
parameters such that the corresponding master equation has 
a unique pure steady state. 
On the other hand, the condition (ii) explicitly contains $V_s$; 
this means that we can characterize the structure of a dissipative 
system such that a desired pure state with covariance matrix 
$V_s$ is generated by that dissipative process. 
Later we provide a modification of the condition (ii) that can 
be more suitably used to find such a dissipative system.


\subsection{Examples}

We here give two examples to explain how the theorem is used.


{\it Example 1: Single OPO}. 
Let us first study an ideal optical parametric oscillator (OPO), 
which couples with a vacuum field through one of the end-mirrors. 
The Hamiltonian is described in terms of the annihilation operator 
$\hat a = (\hat q + i \hat p)/\sqrt{2}$ and the creation 
operator $\hat a^\dagger = (\hat q - i \hat p)/\sqrt{2}$ 
as $\hat H = i(\epsilon \hat a^{\dagger 2} - \epsilon^* \hat a^2)/4$, 
where $\epsilon$ denotes effective complex pump intensity proportional 
to $\chi^{(2)}$ coefficient of the nonlinear crystal. 
Also the coupling operator is given by $\hat L=\sqrt{\kappa} \hat a$ 
with $\kappa$ the damping rate of the cavity. 
The corresponding matrices $(G,C)$ then take the following form:
\begin{equation*}
   G=\left(\begin{array}{cc}
        \Re(\epsilon) & \Im(\epsilon) \\
        \Im(\epsilon) & -\Re(\epsilon)
     \end{array}\right),~~
   C=\sqrt{\frac{\kappa}{2}}~(1,~i). 
\end{equation*}
From Theorem 1, the steady state of this system becomes pure if 
and only if 
\begin{equation*}
    K\Sigma C^\top 
     = \left( \begin{array}{c} 
         C\Sigma C^\top \\ 
         C\Sigma G \Sigma C^\top 
       \end{array} \right)
     = \left( \begin{array}{c} 
         0 \\ 
         \kappa\epsilon
       \end{array} \right)
     = 0. 
\end{equation*}
That is, the dissipative process brought about by the coupling 
to the outer vacuum field must introduce decoherence to the 
intra-cavity state as long as it is squeezed ($\epsilon\neq 0$). 
Actually, when $\epsilon=0$ the steady covariance matrix 
\eqref{purev} turns out to be $I_2/2$. 
In conclusion, a single-mode intra-cavity state can become 
pure only when it is the trivial vacuum (or coherent) state. 
\hfill  $\blacksquare$


{\it Example 2: Cascaded OPOs}. 
We next consider the two-mode OPOs shown in Fig. 1 (a), where 
the OPOs are connected through a unidirectional optical field. 
This kind of cascaded system plays an important role in building 
a quantum information network; e.g., entanglement 
distribution was discussed in \cite{Cirac1}. 
The physical setup of each OPO is the same as before, i.e., 
$\hat H_j = i\epsilon_j(\hat a_j^{\dagger 2} - \hat a_j^2)/4$ and 
$\hat L_j = \sqrt{\kappa} \hat a_j$ $(j=1,2)$ with $\hat a_1$ and 
$\hat a_2$ each cavity modes. 
For simplicity we here set the squeezing effectiveness of each 
cavity to be real; $\epsilon_j\in{\bf R}$. 
From the theory of cascaded systems \cite{Charmichael,GoughJamesTAC}, 
the Hamiltonian and the coupling operator of the whole two-mode 
Gaussian system are respectively given by 
\[
   \hat H =\hat H_1 + \hat H_2 
             + (\hat L_2^\dagger \hat L_1 
               - \hat L_1^\dagger \hat L_2)/2i,~~
   \hat L = \hat L_1 + \hat L_2, 
\]
hence the corresponding system matrices read 
\[
    G=\frac{1}{2} 
         \left(\begin{array}{cc|cc}
             0 & 0 & \epsilon_1 & -\kappa \\
             0 & 0 & \kappa & \epsilon_2  \\ \hline
             \epsilon_1 & \kappa & 0 & 0 \\
             -\kappa & \epsilon_2 & 0 & 0
         \end{array} \right),~~
    C=\sqrt{\frac{\kappa}{2}}(1,~1,~i,~i). 
\]
Then $A=\Sigma(G+\Im(C^\dagger C))$ has eigenvalues 
$-(\kappa \pm \epsilon _j)/2$, thus it is Hurwitz when 
$|\epsilon_j| < \kappa~(j=1,2)$. 
Equivalently, when both the OPOs are below threshold, the steady 
state is unique. 
Then from the condition (iii) of Theorem~1 the steady state is 
pure if and only if 
\[
    K\Sigma C^\top 
     = \left( \begin{array}{c} 
         C\Sigma C^\top \\ 
         C\Sigma G \Sigma C^\top \\ 
         C (\Sigma G)^2 \Sigma C^\top \\ 
         C (\Sigma G)^3 \Sigma C^\top 
       \end{array} \right)
     = \frac{i\kappa(\epsilon_1+\epsilon_2)}{4}
       \left( \begin{array}{c} 
         0 \\ 
         4 \\ 
         0 \\ 
         \lambda 
       \end{array} \right)
     =0, 
\]
where 
$\lambda
=\epsilon_1^2+\epsilon_2^2-\epsilon_1\epsilon_2-3\kappa^2\neq 0$. 
Therefore, the system should be engineered to satisfy 
$\epsilon_1+\epsilon_2=0$ to have a unique pure steady state. 
From Eq.~(\ref{purev}) the corresponding covariance matrix is 
given by 
\begin{equation}
\label{cascadev}
   V_{s}
     =\frac{1}{2}
         \left(\begin{array}{cc|cc}
             \kappa/g_- & -\epsilon/g_- & 0 & 0 \\
             -\epsilon/g_- & \kappa/g_- & 0 & 0 \\ 
             \hline
             0 & 0 & \kappa/g_+ & \epsilon/g_+ \\
             0 & 0 & \epsilon/g_+ & \kappa/g_+
         \end{array}\right),
\end{equation}
where $g_{\pm}:=\kappa\pm\epsilon$ and 
$\epsilon:=\epsilon_1=-\epsilon_2$. 
Thus the steady state is a nontrivial entangled pure state other than 
the vacuum when $0<|\epsilon|<\kappa$. 
Actually the {\it logarithmic negativity} \cite{Vidal}, which is 
a convenient computable measure for entanglement, takes a positive 
value; 
\[
     E(V_s) 
       =\frac{1}{2} \mathrm{ln} 
               \frac{\kappa + |\epsilon|}{\kappa -|\epsilon|}>0. 
\]
We again stress that this dissipation-induced entangled state 
is guaranteed to be highly robust. 
That is, any initial state $\hat \rho(0)$ converges into that 
entangled state. 
Such robustness can be clearly observed from Fig.~1~(b) that 
demonstrates the time-evolutions of the fidelity 
$F(t)=\Tr(\hat\rho(t)\hat\rho_s)$ and 
the purity $P(t)=\Tr(\hat\rho(t)^2)$ with several initial states. 
\hfill  $\blacksquare$

\begin{figure}
\includegraphics[width=2.9cm]{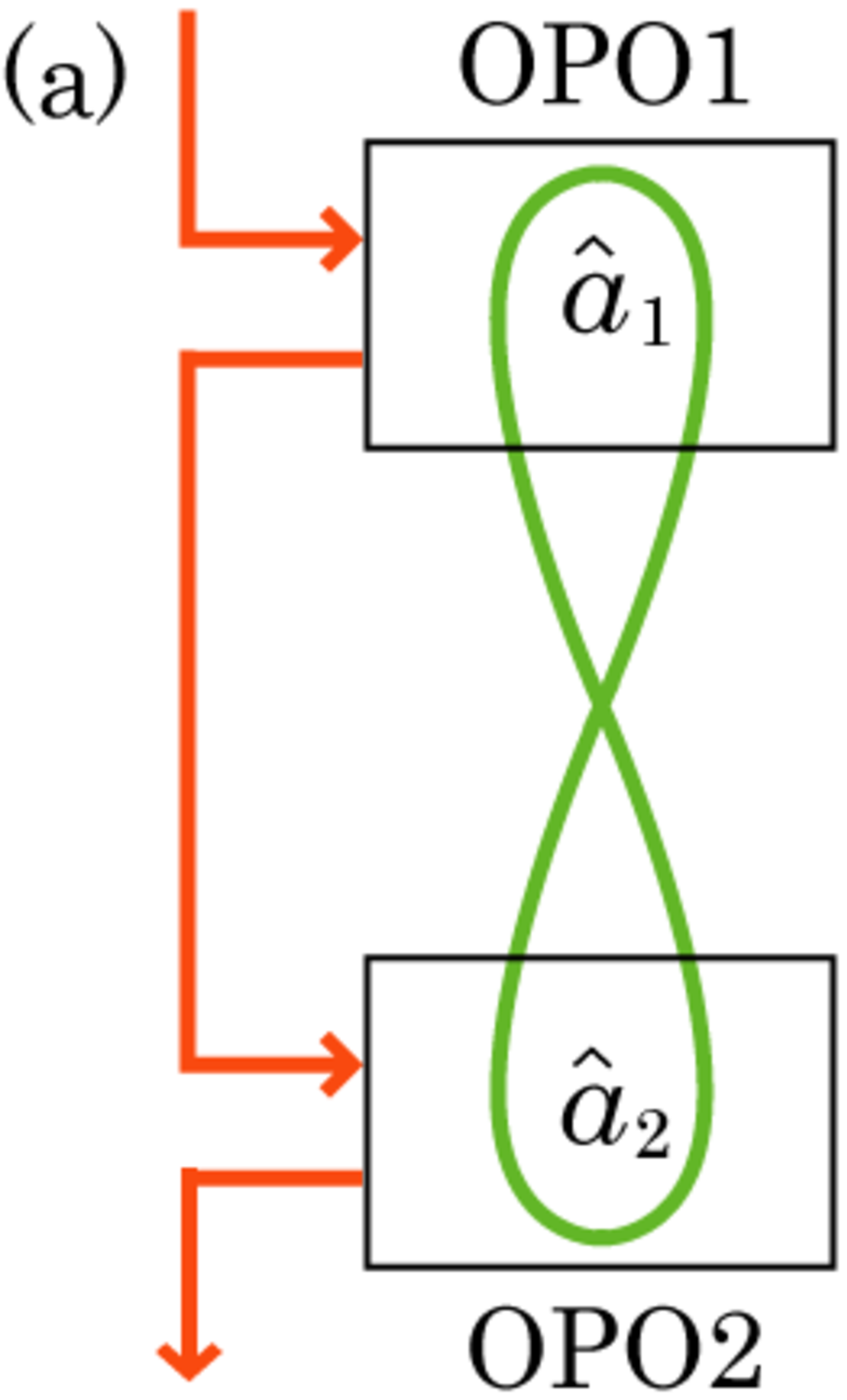}
\includegraphics[width=5.6cm]{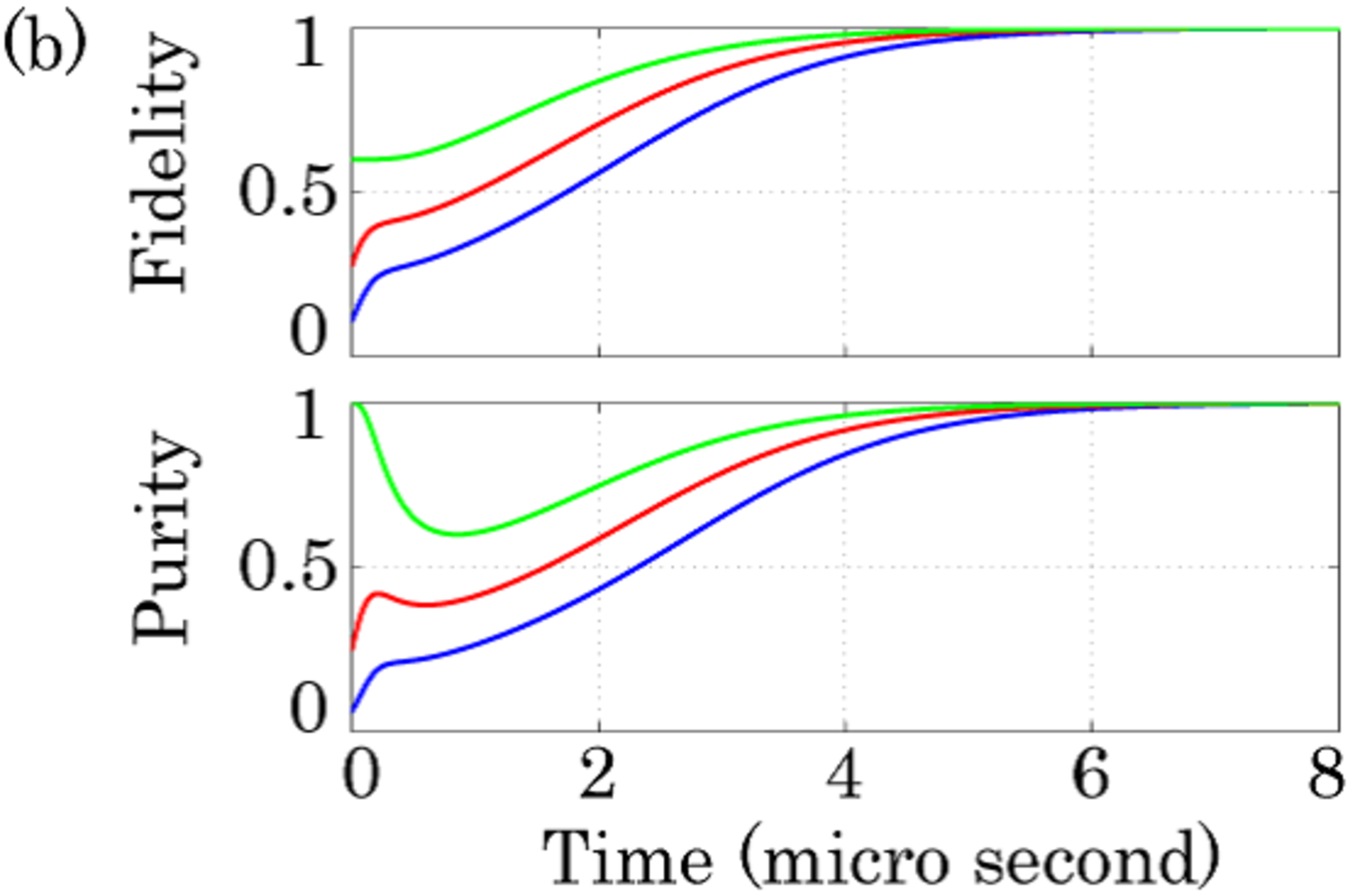}
\caption{
(Color online) 
(a) Schematic diagram of the cascaded OPOs. 
(b) Time evolutions of the fidelity $F(t)$ and the purity $P(t)$ 
of the bipartite cavity state. 
The parameters are $\kappa = 6.0$ MHz and $\epsilon = 4.8$ MHz. 
We take the initial Gaussian states with mean zero and covariance 
matrices $V(0)=I/2$ (top line), $V(0)=I$ (middle), and $V(0)=2I$ 
(bottom). 
Note due to $\mean{\hat x(t)}=0~\forall t$, the fidelity is now 
of the form $F(t)=1/\sqrt{{\rm det}(V(t)+V_s)}$. 
The entanglement achieved in this case is $E(V_s)=1.0986$. 
}
\end{figure}


\section{Environment engineering for pure Gaussian state 
preparation}

In this section we first address a modification of the condition~(ii) 
of Theorem~1 with different assumption, which is more suited to 
the concept of environment engineering. 
In fact, the result allows us to obtain a general procedure for 
synthesizing a dissipative system whose steady state is uniquely 
a desired pure Gaussian state. 
We close this section with some examples.


\subsection{Uniqueness condition for a given Gaussian pure 
steady state}

{\it Theorem 2:} 
Let $V_s$ be a covariance matrix corresponding to a pure Gaussian 
state. 
Then, this is a unique steady state of the system if and only if
\begin{equation}
\label{VtoKc}
   \ker\Big(V_s+\frac{i}{2}\Sigma\Big)= {\rm range}(K^\top), 
\end{equation}
where $K$ is defined in Eq. \eqref{Kc}.

{\it Proof:} 
We begin with the necessary part. 
In general, for a $n$-mode Gaussian pure state, $V_s+i\Sigma/2$ has 
a $n$-dimensional kernel \cite{Simon}, while now from the assumption 
Eq. \eqref{uncertaintyc} holds, implying 
${\rm range}(K^\top)\subseteq\ker(V_s+i\Sigma/2)$. 
Hence, showing ${\rm rank}(K^\top)=n$ completes the proof. 
Now, as shown in the proof of Theorem~1, from the assumption we have 
$K \Sigma K^\top =0$, implying ${\rm rank} (K^\top) \leq n$. 
On the other hand, the uniqueness of the steady state allows us to 
apply Lemma~1 to get ${\rm rank}(\bar{K}^\top) = 2n$, thus 
${\rm rank}(K^\top) \geq n$. 
As a result we obtain ${\rm rank}(K^\top)=n$.

Let us next move to the sufficiency part. 
First we show that $V_s$ satisfying Eq. \eqref{VtoKc} is a solution 
of Eqs. \eqref{th1-1} and \eqref{th1-2}, thus that of Eq.~\eqref{ale}. 
Note that now Eq.~(\ref{th1-1}) apparently holds. 
As seen from the last part of the proof of Theorem~1, Eq.~\eqref{VtoKc} 
implies $(\Sigma G V_s + V_s G \Sigma^\top)K^\top=0$, thus 
we need ${\rm rank}(K'^\top)= 2n$ to derive Eq. \eqref{th1-2}. 
To show this, assume there exists $\xi \in {\bf R}^{2n}$ such that 
$K'\xi=0$. 
This leads to $K\xi=0$. 
Now, from Eq. \eqref{VtoKc} with $V_s$ pure, we have 
$K \Sigma K^\top =0$ and ${\rm rank}(K^\top)=n$, thus 
$\ker(K)$ = ${\rm range} (\Sigma K^\top)$. 
Then, as $\xi \in {\rm range} (\Sigma K^\top)$, we can write 
$\xi=\Sigma K^\top \alpha,~\exists \alpha \in {\bf C}^{2nm}$. 
This leads to 
$0=\Re[(V_s+i\Sigma/2) K^\top \alpha] = V_s\Sigma^\top \xi$, 
thus contradiction to $V_s>0$. 
Consequently, we have ${\rm rank} (K'^\top) =2n$.

Second, to complete the sufficiency part, we show that $V_s$ is 
unique, which is equivalent to that $A$ is Hurwitz. 
Suppose $A^\top\xi = \lambda \xi$ with $\xi \in {\bf C}^{2n}$ and 
$\lambda \in {\bf C}$. 
Multiplying Eq.~\eqref{ale} by $\xi^\dagger$ from the left and by 
$\xi$ from the right, we obtain 
$\Re (\lambda)= -\xi^\dagger D \xi/(2\xi^\dagger V_s \xi)$. 
Note in general $D\geq 0$ and $V_s>0$. 
But now $\xi^\dagger D\xi > 0$, because $\xi^\dagger D\xi = 0$ 
leads to $K'\Sigma ^\top \xi=0$ and this is contradiction to 
${\rm rank}(K'^\top)=2n$. 
Thus $\Re (\lambda)<0$ and this means that $A$ is Hurwitz. 
\hfill  $\blacksquare$

The point of this result is that, while in Theorem~1 the steady 
state is assumed to be unique, we here assume only that a given 
state is pure, without assuming its uniqueness. 
Nonetheless, that state is guaranteed to be a unique steady state 
if the condition \eqref{VtoKc} is satisfied. 
Note again, as seen from the last part of the above proof, that 
the uniqueness is ensured by ${\rm rank}(K^\top)=n$, which leads 
to the Hurwitz property of the matrix $A$. 
That is, we do not need to check whether $A$ is Hurwitz.


\subsection{Complete parameterization of the dissipative system}

Based on the result shown above, we here provide a complete 
parameterization of the Gaussian dissipative system that uniquely 
has a pure steady state. 
This then leads to an explicit procedure for engineering a 
desired Gaussian dissipative system.

We begin with the fact that any covariance matrix $V_s$ 
corresponding to a pure Gaussian state has the following 
general representation \cite{Menicucci2011,Simon88}: 
\begin{equation}
\label{pure CM general}
    V_s=\frac{1}{2}SS^\top,~~~
    S = \left( \begin{array}{cc} 
            Y^{-1/2}  & 0 \\
            XY^{-1/2} & Y^{1/2} \\
        \end{array}\right), 
\end{equation}
where $X$ and $Y$ are $n\times n$ real symmetric and real 
positive definite matrices (i.e., $Y=Y^\top>0$), respectively. 
Note that $S$ is symplectic. 
With this representation we have 
\[
     V_s+\frac{i}{2}\Sigma
       =\frac{1}{2}
        \left( \begin{array}{c} 
            I \\
            Z\dgg \\
        \end{array}\right)
        Y^{-1}(I,~Z), 
\]
where we defined $Z=X+iY$. 
It was shown in \cite{Menicucci2011} that the symmetric matrix 
$Z$ is useful in graphical calculus for several Gaussian pure 
states. 
Because $(I, Z)$ is clearly of rank $n$, we have
\[
    \ker\Big(V_s+\frac{i}{2}\Sigma\Big) 
       = {\rm range}
           \left( \begin{array}{c} 
             -Z \\
             I  \\
           \end{array}\right). 
\]
Hence, Eq. \eqref{VtoKc} is equivalent to
\[
    {\rm range}
           \left( \begin{array}{c} 
             -Z \\
             I  \\
           \end{array}\right)
    ={\rm range}
       (C^\top, G\Sigma^\top C^\top, \ldots, 
          (G\Sigma^\top)^{2n-1}C^\top). 
\]
To satisfy this condition, it is necessary that 
${\rm range}(C^\top)$ is included in ${\rm range}(-Z, I)^\top$ 
and that ${\rm range}(-Z, I)^\top$ is invariant under $G\Sigma^\top$. 
These conditions are respectively represented by 
\begin{equation}
\label{C G condition}
    C^\top=\left( \begin{array}{c} 
             -Z \\
             I  \\
           \end{array}\right)P,~~~
    G\Sigma^\top
           \left( \begin{array}{c} 
             -Z \\
             I  \\
           \end{array}\right)
          =\left( \begin{array}{c} 
             -Z \\
             I  \\
           \end{array}\right)Q, 
\end{equation}
where $P$ and $Q$ are $n\times m$ and $n\times n$ complex 
matrices. 
From the above equations, $K^\top$ is represented by 
\[
   K^\top
      =\left( \begin{array}{c} 
             -Z \\
             I  \\
       \end{array}\right)
         (P,~QP, \ldots,~Q^{2n-1}P). 
\]
Consequently, the necessary and sufficient condition for 
${\rm range}(K^\top)$ to be identical to ${\rm range}(-Z, I)^\top$ 
is that there exist $P$ and $Q$ satisfying Eq. \eqref{C G condition} 
and the rank condition 
\begin{equation}
\label{rank condition}
   {\rm rank}(P,~QP, \ldots,~Q^{n-1}P)=n. 
\end{equation}
Now let us write $G$ in the $2\times 2$ block matrix form 
$G = (G_1, G_2 ; G_2^\top, G_3)$, where $G_1$ and $G_3$ are 
real $n\times n$ symmetric matrices and $G_2$ a real $n\times n$ 
matrix. 
Then the latter equation in Eq. \eqref{C G condition} leads to 
\begin{eqnarray}
& & \hspace*{-1em}
    G_1 + G_2 X + XG_2^\top + XG_3X-YG_3Y=0, 
\nonumber \\ & & \hspace*{-1em}
    (G_2+XG_3)Y+Y(G_2+XG_3)^\top=0. 
\nonumber
\end{eqnarray}
The second equation is equivalent to that there exists a real 
skew symmetric matrix $\Gamma$ (i.e., $\Gamma+\Gamma^\top=0$) 
satisfying $(G_2+XG_3)Y=\Gamma$. 
Hence, by writing $R=G_3$, we find that $G_2$ is expressed as 
$G_2=-XR+\Gamma Y^{-1}$. 
In this representation, $Q$ is of the form 
$Q=-iRY-Y^{-1}\Gamma^\top$. 
To conclude, we obtain the complete parameterization of $C$ and 
$G$ as follows:
\begin{widetext}
\begin{equation}
\label{C G parameterization}
    C=P^\top(-Z,~~I),~~~
    G=\left( \begin{array}{cc} 
          XRX + YRY - \Gamma Y^{-1}X - XY^{-1}\Gamma^\top~~ 
                          & -XR + \Gamma Y^{-1} \\
          -RX + Y^{-1}\Gamma^\top & R        \\
      \end{array}\right). 
\end{equation}
\end{widetext}
Again, $P$ (complex), $R$ (real symmetric), and $\Gamma$ 
(real skew) are the parameter matrices. 
We now have a reasonable procedure for environment engineering 
for pure Gaussian state preparation; 
that is, the procedure is simply to choose the matrices $P, R$, 
and $\Gamma$ so that both the dissipative channels 
$\hat L_k=c_k^\top\hat x~(k=1,\ldots,m)$ and the Hamiltonian 
$\hat H=\hat x^\top G\hat x/2$ with the system matrices given 
in Eq. \eqref{C G parameterization} have desired structures 
such as quasi-locality, while, at the same time, $P$ and 
$Q=-iRY-Y^{-1}\Gamma^\top$ satisfy the rank condition 
\eqref{rank condition}.

Here we give some remarks.

(i) 
There always exists the pair of $(P, R, \Gamma)$ satisfying 
the rank condition Eq. \eqref{rank condition}, if no restriction 
is imposed on those matrices; 
this means that, for any pure Gaussian state, there always exists 
a Gaussian dissipative system for which that state is the unique 
steady state.

(ii) 
The most simple system may be such that $P$ is of rank $n$. 
In this case we can set $R=\Gamma=0$, implying that the system 
does not need a nontrivial Hamiltonian but drives the state 
only by dissipation. 
This kind of system is called the {\it purely dissipative system}. 
However, we are often in the situation where only $m<n$ dissipative 
channels can be implemented in reality, due to some reasons related 
to physical constraints. 
In this case, the matrix $Q$ needs to be of rank at least $n-m$, 
meaning that we must add a nontrivial Hamiltonian.

(iii) 
As stated in \cite{Kraus}, quasi-locality is indeed essential 
since otherwise it would be experimentally hard to realize such 
a dissipative system. 
It should be noticed that, because $\hat H$ is quadratic, it 
can be always decomposed into the sum of quasi-local Hamiltonians 
acting on at most two nodes, although those interactions between 
the nodes do not necessarily have the structure of a target 
entangled state; 
in fact, it will be shown in Example~5 that, to generate 
a chain-type cluster state, a Hamiltonian having a ring-type 
interaction is added. 
That is, while in Gaussian case the quasi-locality issue 
appears only in the part of dissipative channel, this does not 
mean that the complementary Hamiltonian can readily be 
implemented.


\subsection{Examples}

{\it Example 3: General CV cluster and ${\cal H}$-graph states}. 
Menicucci et al. developed in \cite{Menicucci2011} a unified 
graphical calculus for all pure Gaussian states in terms of 
the matrix $Z=X+iY$. 
One of the important results is that the so-called canonical CV 
cluster state, which can be generated by first squeezing the 
momentum quadrature of all modes and then applying the controlled 
$Z$ operations to the modes according to the graph of the cluster, 
can be generally represented by 
\begin{equation}
\label{CV cluster}
   Z=X+ie^{-2r}I, 
\end{equation}
where $X$ corresponds to the symmetric adjacency matrix 
representing the graph structure of the cluster state and $r$ 
the squeezing parameter. 
For this state, for instance setting $P=I$ and $Q=0$ in Eq. 
\eqref{C G condition} gives a desired purely dissipative system 
with channels 
\[
    \left( \begin{array}{c}
           \hat L_1 \\
           \vdots \\
           \hat L_n \\ 
       \end{array}\right)
         =(-X-ie^{-2r}I, I)\hat x, 
\]
which have the same structure as that of the target cluster state. 
Therefore, if each node of the graph is connected to at most 
three adjacency nodes, each dissipative channel acts on at 
most three modes too, i.e., it is quasi-local.

Another important state discussed in \cite{Menicucci2011} is 
the ${\cal H}$-graph state; 
this state is generated by applying the unitary transformation 
$\hat U={\rm exp}(-i\hat {\cal H}t/\hbar)$ with Hamiltonian 
\begin{equation}
\label{H graph Hamiltonian}
   \hat {\cal H}
     =i\hbar \kappa \sum_{j,k}W_{jk}
                (\hat a_j\dgg\hat a_k\dgg - \hat a_j \hat a_k)
\end{equation}
to the vacuum states $\ket{0}^{\otimes n}$, where $W=(W_{jk})$ 
is the real symmetric matrix representing the graph. 
Note that $\hat {\cal H}$ is the sum of the two-mode squeezing 
Hamiltonians. 
The corresponding covariance matrix is then given by Eq. 
\eqref{pure CM general} with $X=0$ and $Y=e^{-2\alpha W}$, 
hence 
\begin{equation}
\label{H graph}
   Z=ie^{-2\alpha W}, 
\end{equation}
where $\alpha=2\kappa t$. 
Unlike the CV cluster state representation \eqref{CV cluster}, 
$Z$ does not necessarily reflect the graph structure of the state. 
However, for instance when $W$ is self-inverse, i.e., $W^2=I$, 
we have $Z=i\cosh(2\alpha)I-i\sinh(2\alpha)W$. 
Thus, in this case choosing $P=I$ gives a desired purely 
dissipative system acting on the nodes in the same manner as 
the Hamiltonian \eqref{H graph Hamiltonian}. 
\hfill $\blacksquare$


{\it Example 4: Two-mode squeezed state}. 
In the Gaussian formulation we are often interested in the 
two-mode squeezed state, as it approximates the so-called EPR state. 
This is represented as a ${\cal H}$-graph state with the 
Hamiltonian \eqref{H graph Hamiltonian} given by 
\[
     W=\left( \begin{array}{cc}
           0 & 1 \\
           1 & 0 \\ 
       \end{array}\right)~~\leftrightarrows~~
     \hat {\cal H}
     =2i\hbar \kappa (\hat a_1\dgg\hat a_2\dgg - \hat a_1 \hat a_2). 
\]
Then the system matrices are $X=0$ and 
\[
    Y=e^{-2\alpha W}
     =\left( \begin{array}{cc}
           \cosh(2\alpha) & -\sinh(2\alpha) \\
           -\sinh(2\alpha) & \cosh(2\alpha) \\ 
      \end{array}\right). 
\]
The corresponding covariance matrix \eqref{pure CM general} is 
given by 
\[
    V_s = \frac{1}{2} 
             \left( \begin{array}{cc|cc} 
                \cosh(2\alpha)  & \sinh(2\alpha) & 0 & 0 \\
                \sinh(2\alpha) & \cosh(2\alpha)  & 0 & 0 \\ \hline
                0 & 0 & \cosh(2\alpha) & -\sinh(2\alpha) \\
                0 & 0 & -\sinh(2\alpha) & \cosh(2\alpha) 
             \end{array}\right). 
\]

Let us begin with constructing a purely dissipative system 
whose unique steady state is the above two-mode squeezed state; 
in this case, we only need to specify a $2\times 2$ full-rank 
matrix $P$ to satisfy the rank condition \eqref{rank condition}. 
In particular, let us here choose 
\begin{equation}
\label{P matrix in example 4}
    P=\left( \begin{array}{cc}
           i\cosh(\alpha) & i\sinh(\alpha) \\
           i\sinh(\alpha) & i\cosh(\alpha) \\ 
      \end{array}\right), 
\end{equation}
which is clearly of full rank. 
Then, $C=P^\top(-Z,I)=(c_1, c_2)^\top$ is given by
\begin{equation}
\label{c1 and c2}
   c_1 = (\mu,~\nu,~i\mu,~-i\nu)^\top,~~
   c_2 = (\nu,~\mu,~-i\nu,~i\mu)^\top, 
\end{equation}
where $\mu=\cosh(\alpha)$ and $\nu=-\sinh(\alpha)$. 
As a result, the master equation describing this purely 
dissipative process is given by 
\begin{equation}
    \frac{d\hat\rho}{dt} 
      = \sum_{k=1,2}\Big(
           \hat L_k\hat \rho \hat L_k^\dagger 
              - \frac{1}{2}\hat L_k^\dagger \hat L_k \hat \rho 
              - \frac{1}{2}\hat \rho \hat L_k^\dagger \hat L_k \Big), 
\end{equation}
where $\hat L_1 = \mu \hat a_1 + \nu \hat a_2^\dagger$ and 
$\hat L_2 = \mu \hat a_2 + \nu \hat a_1^\dagger$. 
A possible physical realization of this system in a pair of 
atomic ensembles was discussed in \cite{Cirac2,Polzik}.

Next let us discuss the case where we are allowed to implement 
only one dissipative channel. 
As an example we take $P=(i\cosh(\alpha), i\sinh(\alpha))^\top$, 
which corresponds to the first column vector of Eq. 
\eqref{P matrix in example 4}. 
This yields $C=c_1^\top$ in Eq. \eqref{c1 and c2}, thus the 
dissipative channel is 
$\hat L_1 = \mu \hat a_1 + \nu \hat a_2^\dagger$. 
We now need to specify a valid $Q$ to satisfy the rank condition 
\eqref{rank condition}; 
in particular let us take a specific Hamiltonian matrix 
$G$ in Eq. \eqref{C G parameterization} with $R={\rm diag}\{0,~1\}$ 
and $\Gamma=0$, then this leads to 
\[
    Q=-iRY-Y^{-1}\Gamma^\top
          =\left( \begin{array}{cc}
             0 & 0 \\
            i\sinh(2\alpha) & -i\cosh(2\alpha) \\ 
           \end{array}\right). 
\]
It is easily verified that $(P,~QP)$ is of full rank, hence the 
requirement is satisfied. 
In this case the Hamiltonian $\hat H=\hat x^\top G\hat x/2$ 
is given by 
\[
   \hat H=(\hat q_1 \sinh(2\alpha)-\hat q_2 \cosh(2\alpha))^2 
             + \hat p_2^2. 
\]
This Hamiltonian and the dissipative channel 
$\hat L= \mu \hat a_1 + \nu \hat a_2^\dagger$ construct 
the desired dissipative system. 
\hfill $\blacksquare$


{\it Example 5: 1-dimensional harmonic chain}. 
As a typical cluster state let us take a 1-dimensional 
(equally weighted) harmonic chain, particularly in the case 
of four-mode cluster just for simplicity. 
Within the formalism of the canonical CV cluster state generation, 
the adjacency matrix $X$ and the graph matrix \eqref{CV cluster} 
are respectively given by 
\[
    X=\left( \begin{array}{cccc}
            0 & 1 &   &  \\
            1 & 0 & 1 &   \\
              & 1 & 0 & 1 \\
              &   & 1 & 0 \\ 
        \end{array}\right),~~
    Z=\left( \begin{array}{cccc}
           ie^{-2r} & 1 & &     \\
           1 & ie^{-2r} & 1 &   \\
             & 1 & ie^{-2r} & 1 \\
             &   & 1 & ie^{-2r} \\ 
        \end{array}\right). 
\]
A desired purely dissipative system is readily obtained 
by taking $P=I_4$ in Eq. \eqref{C G condition}, i.e., 
$C=(-Z,~I)$; 
this means that the four dissipative channels are given by 
\begin{eqnarray}
\label{dissipative channels for chain}
& & \hspace*{-1em}
   \hat L_1 = (-ie^{-2r}\hat q_1+\hat p_1)
                -\hat q_2,
\nonumber \\ & & \hspace*{-1em}
   \hat L_2 = -\hat q_1 + (-ie^{-2r}\hat q_2+\hat p_2) -\hat q_3,
\nonumber \\ & & \hspace*{-1em}
   \hat L_3 = -\hat q_2 + (-ie^{-2r}\hat q_3+\hat p_3) -\hat q_4,
\nonumber \\ & & \hspace*{-1em}
   \hat L_4 = -\hat q_3 + (-ie^{-2r}\hat q_4+\hat p_4). 
\end{eqnarray}
Each channel acts on at most three nodes, thus they are 
quasi-local; 
Fig. 2 (a) depicts the structure of this environment-system 
interaction. 
Note from the above discussion that the general 1-dimensional 
harmonic chain can also be generated by purely dissipative 
process with their channels acting on at most three nodes. 
The finite dimensional counterpart to this result is found 
in \cite{Kraus}.

Now we have a natural question; 
can the chain state be generated by a quasi-local dissipative 
process acting on at most {\it two} adjacency nodes? 
If this is true, this means that engineering the dissipative 
environment becomes easier, apart from that we clearly need 
an additional Hamiltonian. 
Again let us consider the case of four-mode chain and take 
$P=(1, 0, 0, 0)^\top$, implying that the system has one 
dissipative channel $\hat L_1$ in Eq. 
\eqref{dissipative channels for chain}. 
Note this acts on only two nodes. 
To determine the Hamiltonian we have some freedom, 
but let us take in Eq. \eqref{C G parameterization}
\[
   R=X^{-1}=\left( \begin{array}{cccc}
               0 & 1 & 0 & -1 \\
               1 & 0 & 0 & 0  \\
               0 & 0 & 0 & 1  \\
               -1 & 0 & 1 & 0 \\ 
            \end{array}\right),~~
   \Gamma=0. 
\]
This leads to $Q=-iRY-Y^{-1}\Gamma^\top=-ie^{-2r}X^{-1}$, and 
it is readily verified that $(P, QP, Q^2P, Q^3P)$ is of full rank. 
Hence, we obtain the positive answer to the question raised above. 
The matrix $G$ is now of the form 
\[
    G=\left( \begin{array}{cc}
        X+e^{-2r}X^{-1} & -I     \\
        -I              & X^{-1} \\ 
            \end{array}\right). 
\]
The first $(1,1)$ block matrix indicates that the Hamiltonian 
$\hat H=\hat x^\top G\hat x/2$ has a ring-type structure where 
the 1-2, 2-3, 3-4, and 4-1 nodes are connected with each other; 
see the remark (iii) in the previous subsection. 
Therefore it is concluded that the four-mode harmonic chain 
state can be generated by the dissipative process where both 
the dissipative channel and the Hamiltonian quasi-locally act 
on only two adjacency nodes; 
the structure of these interactions is shown in Fig. 2 (b). 
Implementation of this system should be easier than that of 
the purely dissipative system obtained above, which acts on 
the nodes through the channels \eqref{dissipative channels for 
chain}. 
\hfill $\blacksquare$

\begin{figure}[htbp]
\begin{center}
\includegraphics[width=7.4cm]{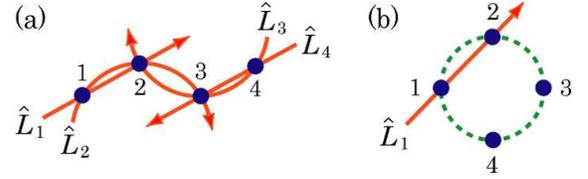}
\caption{
(Color online) 
(a) Dissipative channels for generating the harmonic chain state. 
(b) A combination of the dissipative channel (arrow) and the 
ring-type Hamiltonian (dotted circle) can generate the harmonic 
chain state. 
}
\end{center}
\end{figure}


\section{Conclusion}

In this paper, we have derived the general necessary and sufficient 
conditions for a Gaussian dissipative system to have a unique pure 
steady state. 
In particular, we have provided a complete parameterization of the 
Gaussian dissipative system satisfying the requirement; 
this leads to an explicit procedure for engineering a Gaussian 
dissipative system whose steady state is uniquely a desired pure 
state.

An important open question is how to construct, in a general 
framework, a practical dissipative system satisfying a quasi-locality 
constraint. 
Although Example~5 demonstrated the feasibility in engineering 
such a quasi-local dissipative system, that construction method 
is a heuristic one. 
Note that in the finite-dimensional case the same problem 
remains open \cite{Kraus}.

We can take two approaches to dealing with the problem. 
The first one is suggested by the recent work by Ticozzi and Viola 
\cite{TicozziViola2011} that, in the finite-dimensional case, 
characterizes a pure state generated by a dissipative system having 
a fixed quasi-locality constraint; 
it was then shown that, based on this result, by switching 
dissipative channels through output feedback control \cite{Wiseman}, 
we are enabled to construct a desired quasi-local dissipative system. 
It is expected that this idea works in our case as well. 
Actually, it was shown in \cite{Ficek2009} that, with the use of 
a similar switching method, some Gaussian cluster states can be 
generated in four-mode atomic ensembles trapped in a ring cavity.

The other approach uses the schematic of quantum state transfer. 
The basic idea is as follows. 
First, an $n$-mode target (entangled) Gaussian state of light fields 
is produced, and then, those fields are independently coupled to 
$n$ identical bosonic systems in a dissipative way; 
then, as a result, that state is deterministically transferred 
to the systems. 
In this scheme, we need the strong assumption that all the nodes 
are accessible and independently couple to the fields, but the 
interactions between the systems and the environment channels 
are all {\it local}. 
Hence, from the fact that a number of pure Gaussian cluster state 
of light fields can be produced efficiently with the use of some 
beam splitters and OPOs \cite{Zhang2006,vanLoock2007,Menicucci2011}, 
this state-transfer-based approach is expected to be more reasonable. 
In fact, the dissipation-induced states shown in \cite{Cirac1,
Cirac2,Polzik} are generated using this technique. 
Furthermore, one of the authors recently has developed a general 
theory of this scheme in terms of a quantum stochastic differential 
equation \cite{YamamotoRS2011}.

It would be worth to further examine the above two approaches 
to make them more useful for engineering a practical dissipative 
system.





\end{document}